\documentclass[twocolumn,prl,longbibliography,superscriptaddress,floatfix,footinbib]{revtex4-1}

\usepackage[dvipsnames]{xcolor}
\usepackage{graphicx}
\usepackage{amssymb}
\usepackage{amsmath}
\usepackage{amsfonts}
\usepackage{mathtools}
\colorlet{Mycolor1}{Plum}
\usepackage{tikz}
\usepackage[normalem]{ulem}

\newcommand{\fig}[1]{Fig.~\ref{#1}}

\newcommand{\eq}[1]{Eq.~\ref{#1}}

\makeatletter
\newsavebox{\@brx}
\newcommand{\llangle}[1][]{\savebox{\@brx}{\(\m@th{#1\langle}\)}%
  \mathopen{\copy\@brx\kern-0.5\wd\@brx\usebox{\@brx}}}
\newcommand{\rrangle}[1][]{\savebox{\@brx}{\(\m@th{#1\rangle}\)}%
  \mathclose{\copy\@brx\kern-0.5\wd\@brx\usebox{\@brx}}}
\makeatother

\usepackage[caption=false]{subfig}
\usepackage{graphicx}

\usepackage{bm}

\newcommand{\be}{\begin{equation}}
\newcommand{\ee}{\end{equation}}

\newcommand{\bea}{\begin{equation}\begin{aligned}}
\newcommand{\eea}{\end{aligned}\end{equation}}

\usepackage[colorlinks=true,
            linkcolor=red,
            citecolor=blue,
            urlcolor=blue]{hyperref}

\begin{document}

\def\titlename{Hidden Critical Points in the Two-Dimensional $O(n>2)$ model:\\ Exact Numerical Study of a Complex Conformal Field Theory }
\title{\titlename}

\author{Arijit Haldar}
\thanks{These two authors contributed equally.}
\affiliation{Department of Physics, University of Toronto, 60 St. George Street, Toronto, Ontario, M5S 1A7, Canada}
\affiliation{S. N. Bose National Centre for Basic Sciences, JD Block, Sector-III, Salt Lake City, Kolkata - 700 106, India}
\author{Omid Tavakol}
\thanks{These two authors contributed equally.}
\affiliation{Department of Physics, University of Toronto, 60 St. George Street, Toronto, Ontario, M5S 1A7, Canada}
\affiliation{Department of Physics and Astronomy, University of California, Irvine, California 92697, USA}
\author{Han Ma}
\affiliation{Perimeter Institute for Theoretical Physics,
Waterloo, Ontario N2L 2Y5, Canada}
\author{Thomas Scaffidi}
\affiliation{Department of Physics and Astronomy, University of California, Irvine, California 92697, USA}
\affiliation{Department of Physics, University of Toronto, 60 St. George Street, Toronto, Ontario, M5S 1A7, Canada}

\begin{abstract}
The presence of nearby conformal field theories (CFTs) hidden in the complex plane of the tuning parameter was recently proposed as an elegant explanation for the ubiquity of ``weakly first-order'' transitions in condensed matter and high-energy systems. In this work, we perform an exact microscopic study of such a complex CFT (CCFT) in the two-dimensional $O(n)$ loop model. The well-known absence of symmetry-breaking of the $O(n>2)$ model is understood as arising from the displacement of the non-trivial fixed points into the complex temperature plane.
Thanks to a numerical finite-size study of the transfer matrix, we confirm the presence of a CCFT in the complex plane and extract the real and imaginary parts of the central charge and scaling dimensions. By comparing those with the analytic continuation of predictions from Coulomb gas techniques, we determine the range of validity of the analytic continuation to extend up to $n_g \approx 12.34$, beyond which the CCFT gives way to a gapped state. Finally, we propose a beta function which reproduces the main features of the phase diagram and which suggests an interpretation of the CCFT as a liquid-gas critical point at the end of a first-order transition line.

\end{abstract}

\maketitle
\textit{Introduction}--- 
 The notion of universality at continuous phase transitions is central to our understanding of most phases of matter.  However, there are several examples of ``weakly first-order'' transitions in high-energy and condensed matter physics which appear continuous at intermediate scales but eventually turn out to be first-order at larger scales\cite{(2)_Singularities_Scaling_Functions),(2)_scaling_theory, Cardy1980, Fukugita_1989, Itakura2003, conformal_loss_2009,Adam_2_DQC_classical_loop,Adam_DQCP_duality,Gorbenko2018a,Gorbenko2018b,(1)_weakly_phase_transition,Adam_weakly_first_order,Ma19, Benini2019,Lauchli21}.
 One example is $4D$ gauge theories coupled to matter for which the gauge coupling is conjectured to run slowly (``walking behavior'') at intermediate energies but starts running fast again at low energies, leading to confinement and chiral symmetry breaking (``conformality loss'')\cite{conformal_loss_2009}.
Another example is the 2D classical $Q$-state Potts model, for which the ferromagnetic phase transition is second-order for $Q \leq 4$, but becomes weakly first order for $Q$ slightly above 4\cite{(2)_scaling_theory,(2)_Singularities_Scaling_Functions),Gorbenko2018b,Ma19}.
Further, numerical studies of the transition between Neel and valence bond solid states also point towards a weakly first order scenario\cite{Adam_2_DQC_classical_loop,Adam_weakly_first_order,Adam_DQCP_duality,Lauchli21}.

 Recently, it was proposed that fixed point annihilation~\cite{Adam_2_DQC_classical_loop,Adam_DQCP_duality}, and more specifically the resulting presence of complex conformal field theories (CCFTs) ~\cite{Gorbenko2018a,Gorbenko2018b} hidden in the complex plane of the tuning parameter, could explain the widespread occurrence of weakly first-order transitions. 
In this scenario, the slow RG flow on the real axis is explained by the presence of a nearby CCFT in the complex plane, and the properties of the approximately conformal theory observed at intermediate scales can be derived from the complex conformal data of the CCFT.
Complex CFTs are non-unitary CFTs with highly unusual behavior, since they have complex central charge and scaling dimensions, and the RG flow around them forms a spiral.
 Exploring complex CFTs is also relevant from the perspective of understanding phase transitions in dissipative quantum systems described by non-Hermitian Hamiltonians~\cite{(5)_open_non_hermitain_1,(5)_open_non_hermitain_2,(5)_magnetic_non_hermitian_1,(5)_magnetic_non_hermitian_2,RevModPhys.93.015005,PhysRevX.8.031079,non_hermitian_SPT1,non_hermitian_SPT2,non_hermitian_phase_transition,(6)_non_unitary_1,(6)_non_unitary_2,(6)_non_unitary_3}.

The study of CCFTs is however challenging, and few models and results exist~\cite{Gorbenko2018a,Gorbenko2018b,Faedo2019,Giombi2019,Benini2019,Benedetti_2021,Zan3,Ma19,Faedo2021}.
First, since the tuning parameter needs to be complexified, finding a fixed point requires the tuning of at least two real parameters. 
Second, the study of CCFTs has so far relied on holography~\cite{Faedo2019,Faedo2021}, perturbative methods~\cite{Benini2019}, or on the analytic continuation of real CFTs~\cite{Gorbenko2018b}. 
However, these methods have their limitations: for example, the range of validity of the analytic continuation is not known.

In this work, we propose instead to generalize the non-perturbative numerical methods which exist for 2D CFTs \cite{Blote1989,Batchelor1988,Blote1994} to the case of complex CFTs.
This enables us to provide a complete characterization of microscopic models of 2D CCFTs, including the complex central charge and scaling dimensions, the connection betweenscaling operators and microscopic operators, and the finite-size RG flow.



In order to demonstrate our approach, we work with the $O(n)$ loop model, which is closely related to the $Q$-state Potts model in 2D and has the advantage of being self-tuned to the Potts transition surface~\cite{NIENHUIS_1991}.
The transition surface, residing in the parameter space of the 2D Potts model generalized to include vacancies,  separates the ferromagnetic and paramagnetic phases and contains two fixed points. One of these fixed points belongs to the Potts universality class and the other belongs to the tricritical Potts universality~\cite{NIENHUIS_1991} class. The two fixed points collide and annihilate at $Q=4$, resulting in a weakly first-order transition for $Q \gtrsim 4$ which was recently described in terms of CCFTs in Ref.~\cite{Gorbenko2018b}.
Under the mapping $n=\sqrt{Q}$, a finite (resp. diverging) correlation length in the $O(n)$ loop model corresponds to a first-order (resp. second-order) transition for the $Q$-state Potts model.




In this context, the first-order nature of the transition in the Potts model for $Q>4$ is related to the well-known absence of a symmetry-breaking transition in the $O(n > 2)$ model.
  This absence can be understood as arising from the displacement of $O(n)$ critical points into the complex plane of the $O(n)$ temperature parameter at $n=2$\footnote{Note that the temperature of the $O(n)$ model does not map to the temperature of the Potts model}.
This means the $O(n)$ model actually still harbors a critical point with a diverging correlation length for $n>2$, but for a complex value of the temperature. This critical point is described by a CCFT and should lead to a ``walking'' RG flow on the real axis for $n \gtrsim 2$. We note that a generalization of the CCFT analysis to the $O(n)$ model was already proposed in Refs.~\cite{Gorbenko2018b,Zan3}.


\begin{figure}
    \centering
    \includegraphics[width=0.48\textwidth]{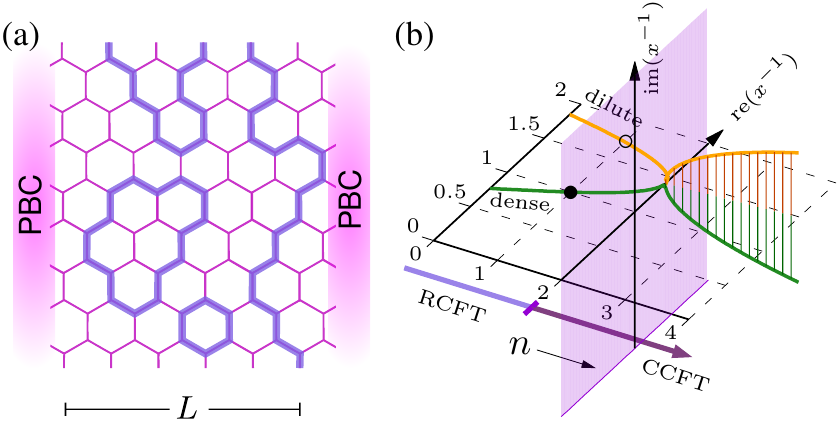}
    \caption{(a) Example of a loop configuration on the hexagonal lattice. (b) The location of the critical branches in the complex $x^{-1}$ plane as a function of $n$. The orange (resp. green) line corresponds to $x_{c,+}$ (resp. $x_{c,-}$). RCFT stands for real CFT.
    }
    \label{fig:model}
\end{figure}

\textit{Model and CFT predictions}---
\label{secII}
Starting from a truncated high-temperature expansion of the $O(n)$ model on the honeycomb lattice, one obtains a model of non-intersecting loops on the same lattice (see Fig. \ref{fig:model}(a)) \cite{Nienhuis82,Peled2017}:
\bea
Z = \sum_{i \in \text{loop config}} n^{N_i} x^{l_i},
\label{Z2}
\eea
where $n$ of the $O(n)$ model is reinterpreted as the loop fugacity and $x^{-1}$ is the loop tension (which corresponds to the temperature of the $O(n)$ model with $x = \beta J$).
For each loop configuration $i$, $N_i$ is the number of loops, and $l_i$ is the total length of all loops.
Note that the loop model is well defined even when $n$ is not an integer.

This model was shown to be critical~\cite{Nienhuis82,Baxter,Blote1989,Batchelor1988} for $-2 \leq n \leq 2$ if the loop tension sits on one of two branches:
\begin{equation}
x=x_{c,\pm}\equiv\left(2\pm(2-n)^{1/2}\right)^{-1/2}\,.
\label{wpm}
\end{equation}
The $x_{c,+}$ branch is the so-called dilute branch and sits at the transition between the short-loop phase in the region $x<x_{c,+}$ (which is equivalent to the high-$T$ paramagnetic phase of the $O(n)$ model) and the critical dense loop phase in the region $x>x_{c,+}$(see Fig.\ref{fig:model}(b)).

Both branches have a CFT description based on Coulomb Gas (CG)~\cite{colomb_gas_1987,Nienhuis82,Baxter,Blote1989,Batchelor1988,Jacobsen2009} techniques with the following central charge:
\begin{align}
    c_\pm(n) =\left(4 - 7e(n)^2 \pm 3 e(n)^3\right) / \left(4 - e(n)^2\right),
    \label{central_charge}
\end{align}
where $e(n) = \left(2/\pi\right)\cos^{-1}{(n/2)} $ is the background charge.
A few notable examples are the Berezinskii–Kosterlitz–Thouless transition with $c_{\pm}(2)=1$, Ising with $c_+(1)=1/2$, percolation with $c_-(1)=0$ and dense polymers with $c_-(0)=-2$.

A number of scaling dimensions are also known, like the thermal and magnetic ones (corresponding in the $O(n)$ notation to the lowest singlet and vector operator, respectively):
\begin{align}
    X_{t\pm} = & 16/g_{\pm}-2 \nonumber\\
    X_{h\pm} = & g_\pm/32 - \left(2/g_\pm\right)\left(1-g_{\pm}/4\right)^2
    \label{scaling}
\end{align}
with $g_\pm(n) =  4\pm 2 e(n)$. The thermal scaling dimension probes the response to a change the loop tension, or equivalently to a change in temperature of the original $O(n)$ model. The magnetic exponent describes the spin-spin correlations of the original $O(n)$ model.

In order to extend the above CFT predictions to the case of complex CFTs for $n>2$, we follow Ref.~\cite{Gorbenko2018b}, in which an analytic continuation of known CFT predictions for the $Q\leq 4$ Potts model was continued to $Q>4$.
Since the $Q$-state Potts model and the $O(n)$ loop models realize the same CFT branches for $Q=n^2$, it is natural to use the same analytic continuation here for the $O(n>2)$ loop model.
(Note however that the operator content is different for these two theories, and that no equivalent of Eq.~\ref{wpm} exists for the Potts model).

\begin{figure}
    \centering
    \includegraphics[width=0.48\textwidth]{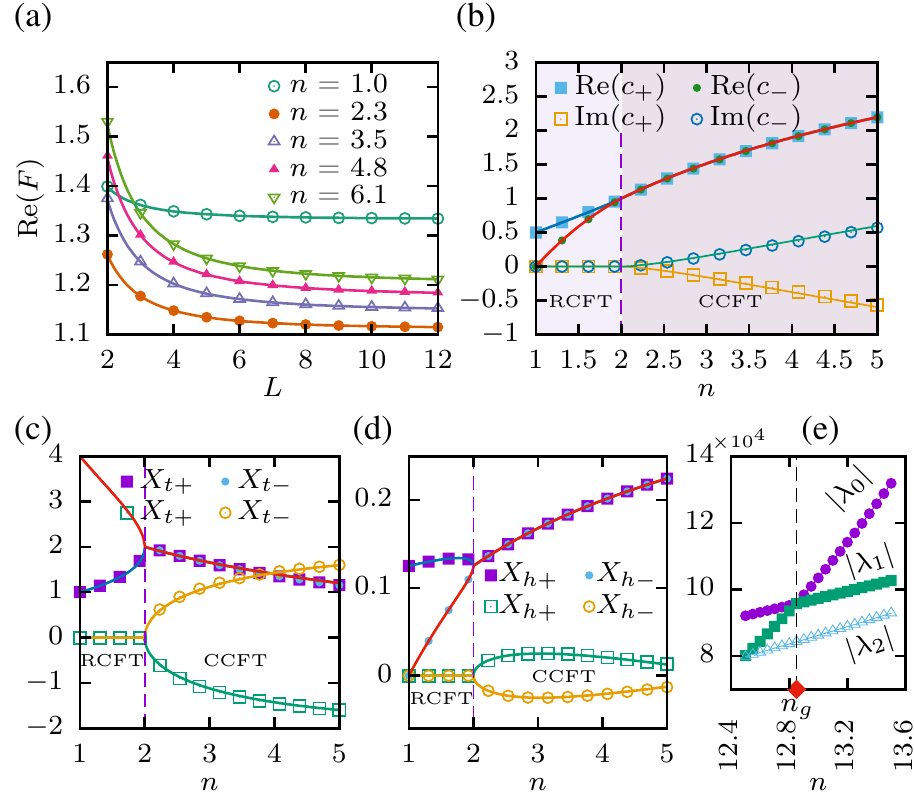}
    \caption{ (a) Real part of the free energy vs system size for various $n$. Solid line shows fit to $F_{L}=F_{\infty}+\frac{\pi c}{6L^2}$. The fits were calculated using numerical data points starting from $L=4$ to $L=12$.
    Real and imaginary parts of (b) the complex central charge $c_\pm$ for the two branches, (c) the thermal scaling dimension $X_{t\pm}$, and (d) the magnetic scaling dimension $X_{h\pm}$.
    In (b)-(d), the solid lines are CFT predictions and the dots are numerical results. All the results are calculated at $x = x_{c,\pm}$. 
   The scaling dimensions $X_{t\pm}$ (c) and $X_{h\pm}$ (d) are obtained for $L=11$.  
   (e) Magnitude of the largest three eigenvalues of the transfer matrix (for $L=10$), showing a transition for $\lambda_0$ at $n=n_g$.}
    \label{fig:results}
\end{figure}

Based on Eq.~\ref{wpm}, one finds that the two critical branches meet at $n=2$, and move to the complex plane for $n>2$ (see Fig. \ref{fig:model}(b)).
Relatedly, one can easily see from Eqs. \ref{central_charge} and \ref{scaling} that the value of the central charge and of the scaling dimensions becomes complex for $n>2$ (see continuous lines in Fig. \ref{fig:results} b-d).
Note that, for $n>2$, the two branches are simply complex conjugates of each other ($x_{c,+} = x_{c,-}^*$, $c_+ = c_-^*$, and $X_+ = X_-^*$), whereas for $n<2$ they correspond to very different physics.

\textit{Numerics at the fixed points}---
We now use transfer matrix numerics to verify the predictions summarized in Eqs~\ref{wpm}, \ref{central_charge}, and \ref{scaling}.
We use periodic boundary conditions (PBCs) along the horizontal direction such that the system forms a long cylinder with a circumference of size $L$ (see Fig.~\ref{fig:model}(a)).
Our implementation of the transfer matrix generalizes the one of Ref.~\cite{Blote1989,BLOTE1989121} and is detailed in the Supplemental Material(SM)~\cite{[{See Supplemental Material at }][]supp}.

If we order the eigenvalues of the transfer matrix by their magnitude, $|\lambda_{L,0}| \geq |\lambda_{L,1}| \geq \dots $, the free energy per site in the long cylinder limit is given by\cite{Blote1989,Batchelor1988,Blote1994}:
\begin{equation}
    F_{L}= \frac{2}{\sqrt{3}L}\log\big(\lambda_{L,0}\big)\,.
    \label{free_energy}
\end{equation}
An estimate for the central charge is then obtained by finite-size scaling through $F_{L}=F_{\infty}+\frac{\pi c}{6L^2} $.

The finite-size estimate of the thermal scaling dimension is related to the gap between $\lambda_{L,0}$ and the subleading eigenvalue $\lambda_{L,1}$:
\begin{equation}
    X_{t}= \frac{2\pi}{L}\log\left(\frac{\lambda_{L,0}}{\lambda_{L,1}}\right)\,.
\end{equation}

In order to calculate the magnetic scaling dimension, it is necessary to define another Hilbert space sector (the so-called magnetic sector) for which there is a single non-contractible loop traversing the whole cylinder vertically. Denoting the leading eigenvalue in that sector as $\tilde\lambda_{L,0} $, the estimate for $X_h$ is:
\begin{equation}
    X_{h}= \frac{2\pi}{L}\log\left(\frac{\lambda_{L,0}}{\tilde\lambda_{L,0}}\right).
\end{equation}

The numerical results for the real and imaginary parts of $c$, $X_t$ and $X_h$ at $x = x_{c,\pm}$ are in perfect agreement with the CFT predictions for a broad range of loop fugacities above $n=2$. We show results up to $n=5$ in Fig.~\ref{fig:results}, but the agreement actually persists until $n=n_g\simeq 12.34$.
We have thus confirmed the existence of CCFTs in the range $n \in [2,n_g]$.

As shown in Fig.~\ref{fig:results}(e) (see also SM~\cite{[{See Supplemental Material at }][]supp}), we find a level crossing at $n=n_g$  \footnote{Note that the two eigenvalues only cross in magnitude but not in phase.} beyond which the transfer matrix is gapped (i.e. $\log(|\lambda_{L,0}|/|\lambda_{L,1}|) \sim \mathcal{O}(1)$), which means the system has a finite correlation length.
By inspection of the dominant eigenstate for $n>n_g$, it appears likely that the corresponding phase is adiabatically connected to the short loop phase obtained for $x \to 0$.
 We leave the study of the range $n>n_g$ for future work and now focus on $n \in [2,n_g]$.

\begin{figure}[t!]
  
  \begin{minipage}{\linewidth}
  \includegraphics[width=.5\linewidth]{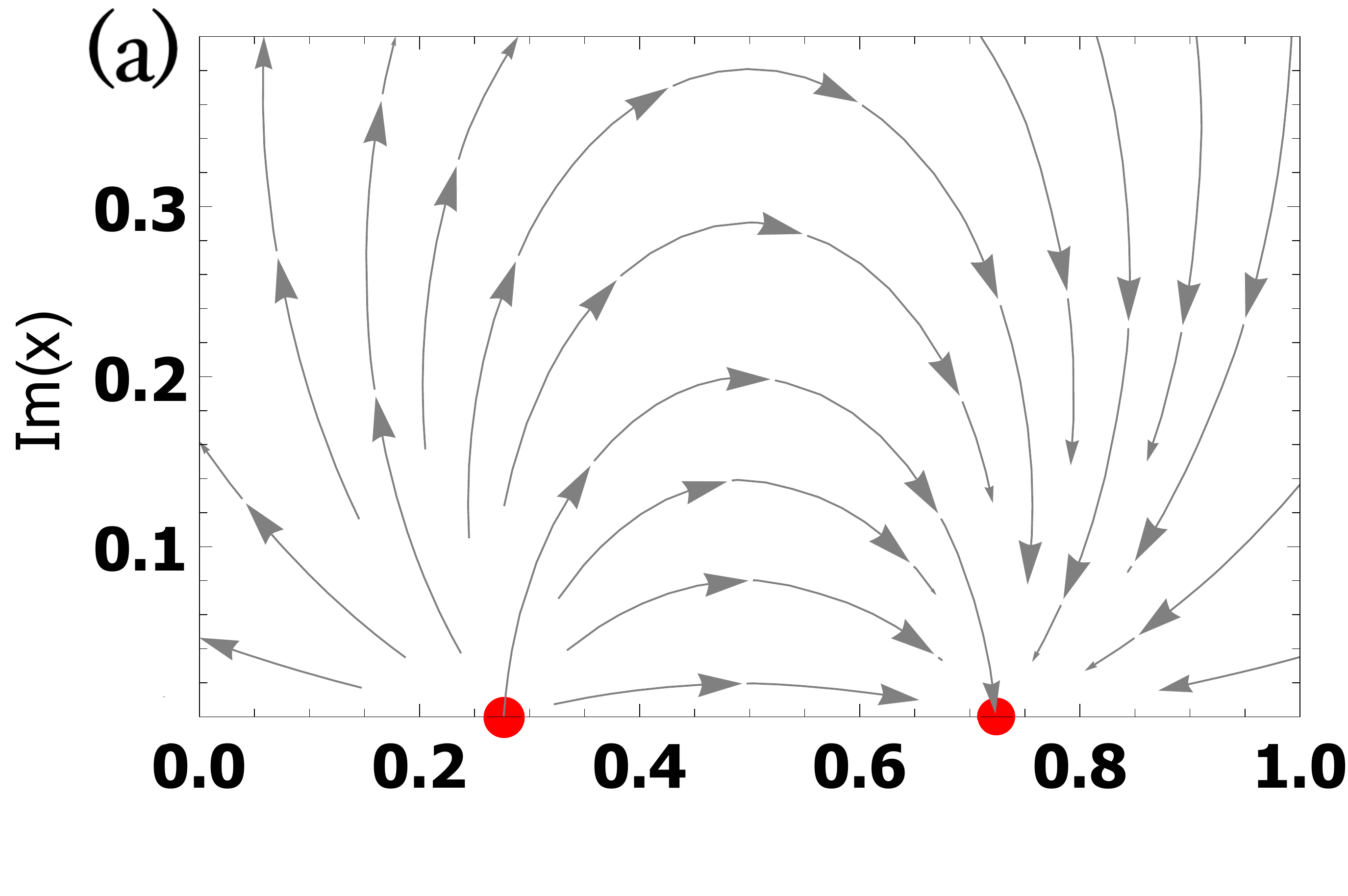}\hfill
  \includegraphics[width=.5\linewidth]{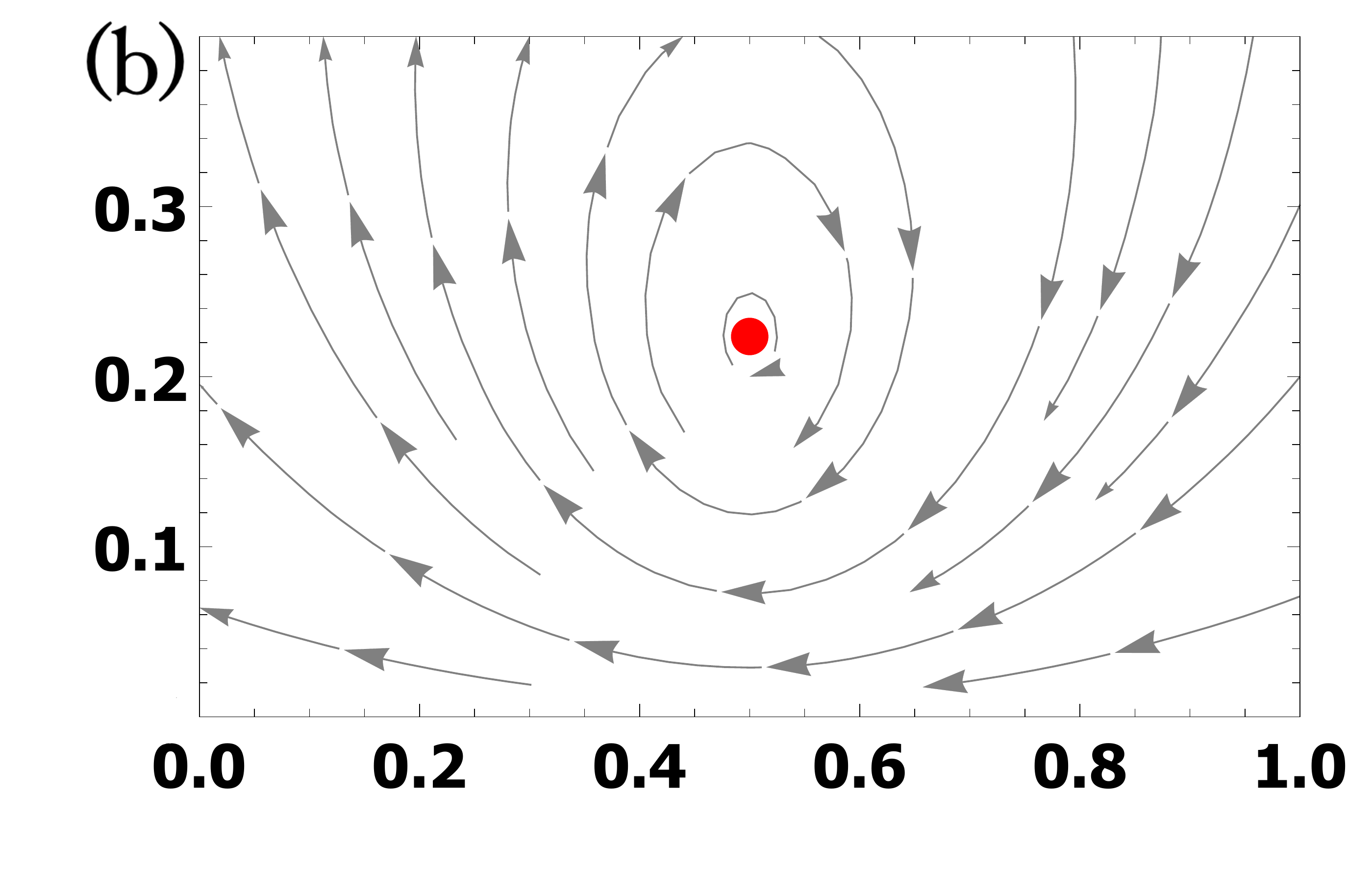}
  \end{minipage}
   \par
  \begin{minipage}{\linewidth}
  \includegraphics[width=0.5\linewidth]{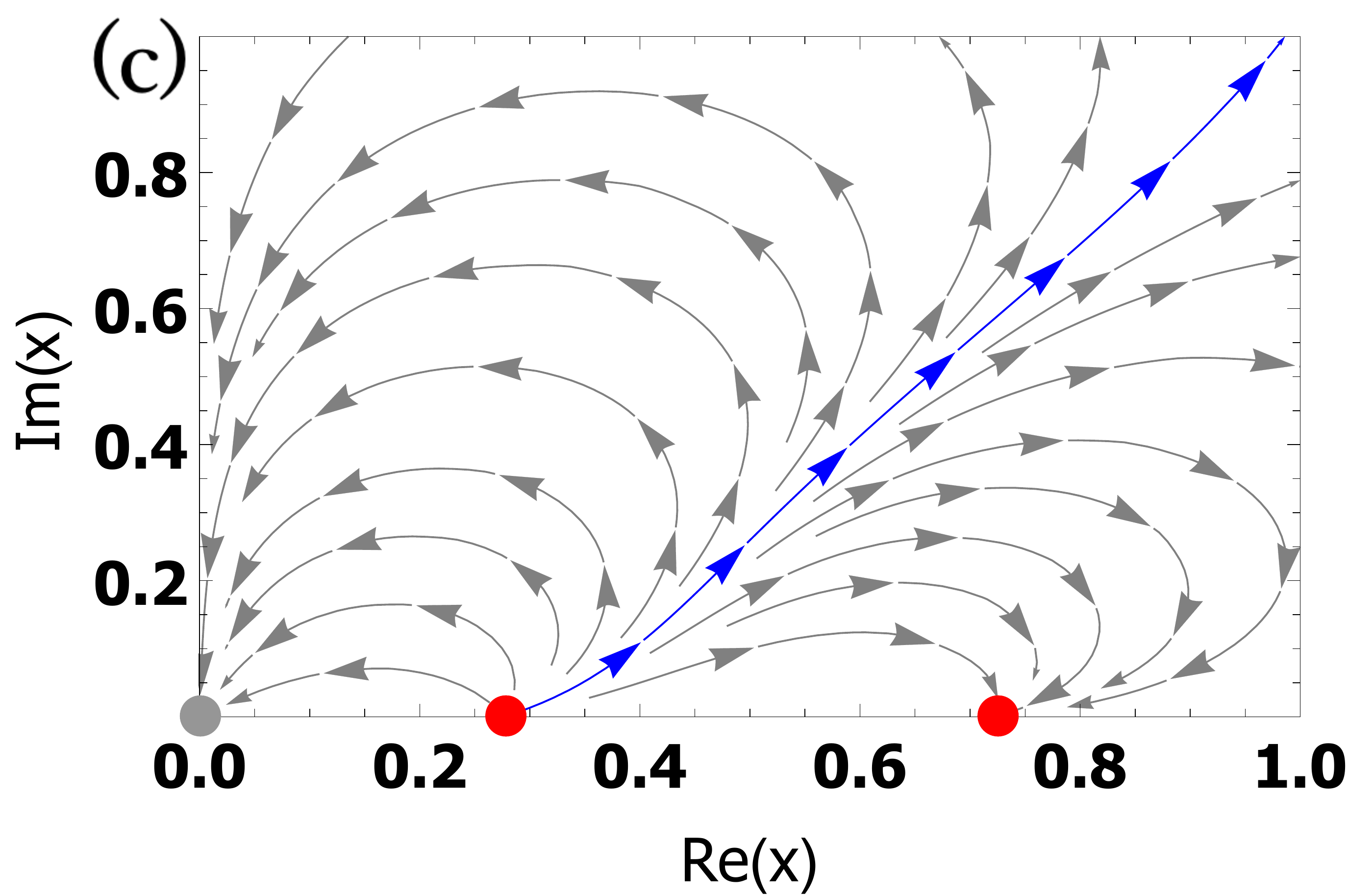}\hfill
  \includegraphics[width=0.5\linewidth]{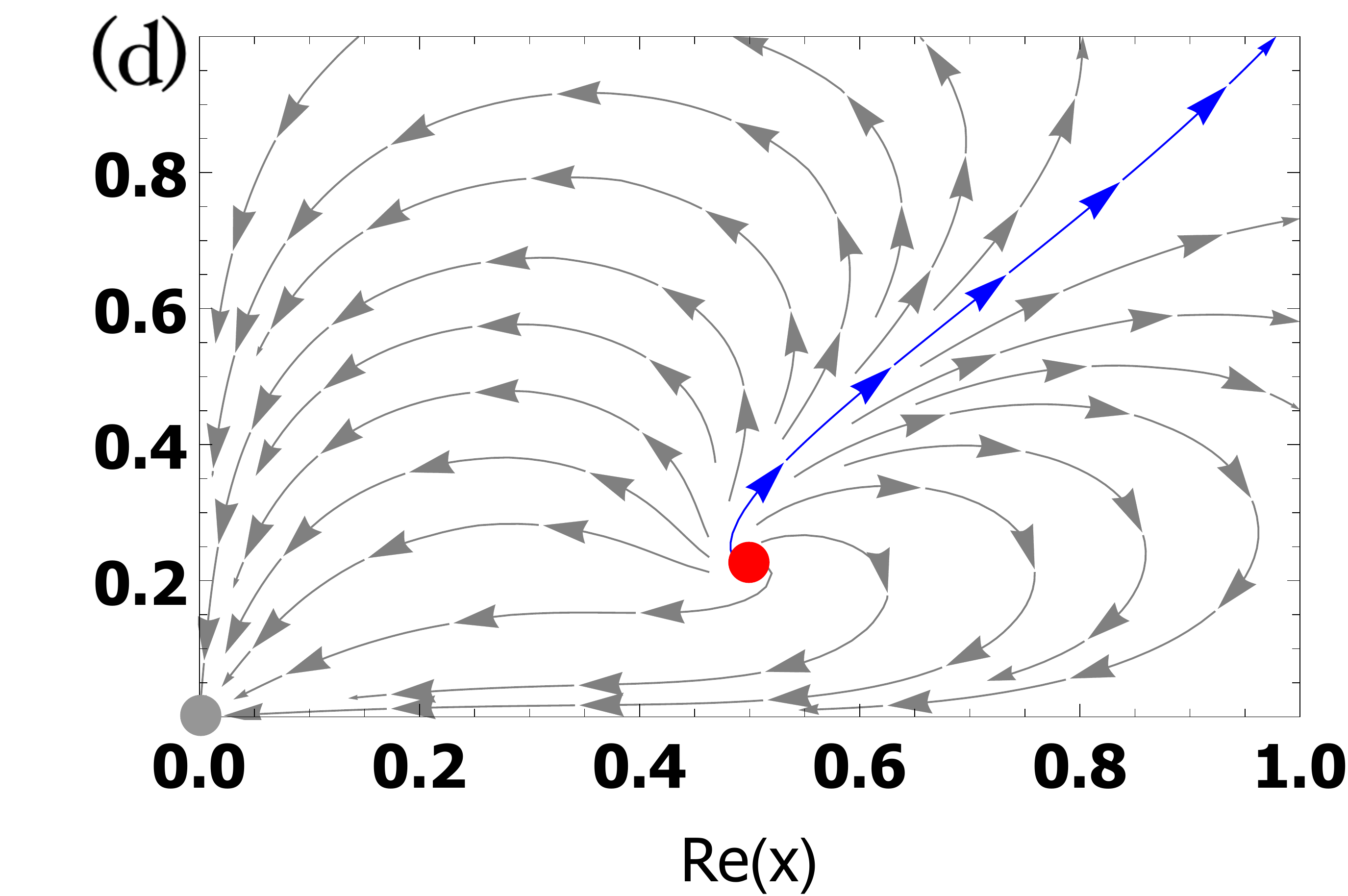} 
  \end{minipage}%
  \par
  \caption{Top: RG flow for the simple beta function  $\beta_\text{sim}(x) = -\mu - (x-x_0)^2$ with $x_0=0.5$ and with (a) $\mu=-0.05$ or (b) $\mu=0.05$. Bottom: RG flow for the generalized beta function $\beta_\text{gen}(x)$ for (c) $\mu=-0.05$ or for (d) $\mu=0.05$. The separatrix is shown in blue. The red dots show the critical points and the grey dot shows the gapped short loop fixed point at $x=0$.}\label{fig:RG1}
\end{figure}

\textit{RG flow and phase diagram}---
Now that we have established the existence of CCFTs for $n>2$, let us discuss their broader significance for the phase diagram and the RG flow.
The standard scenario is that the presence of a complex CFT right above the real axis leads to a slowing down of the RG flow on the real axis, and hence the walking behavior.
This is usually understood with the following ``simple'' beta function: $\beta_\text{sim}(x) = -\mu - (x-x_0)^2$, where $\mu<0$ corresponds to real CFTs on the real axis, and $\mu>0$ corresponds to complex CFTs located at $x=x_0 \pm i \sqrt{\mu}$ (see Figs. \ref{fig:RG1}(a) and (b)).

\begin{figure}[t!]%
\includegraphics[width=0.45\textwidth]{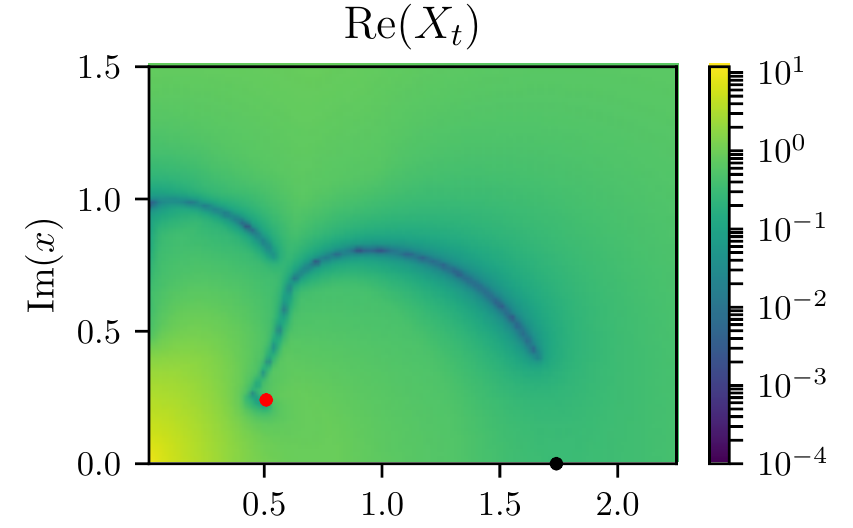} %
\includegraphics[width=0.45\textwidth]{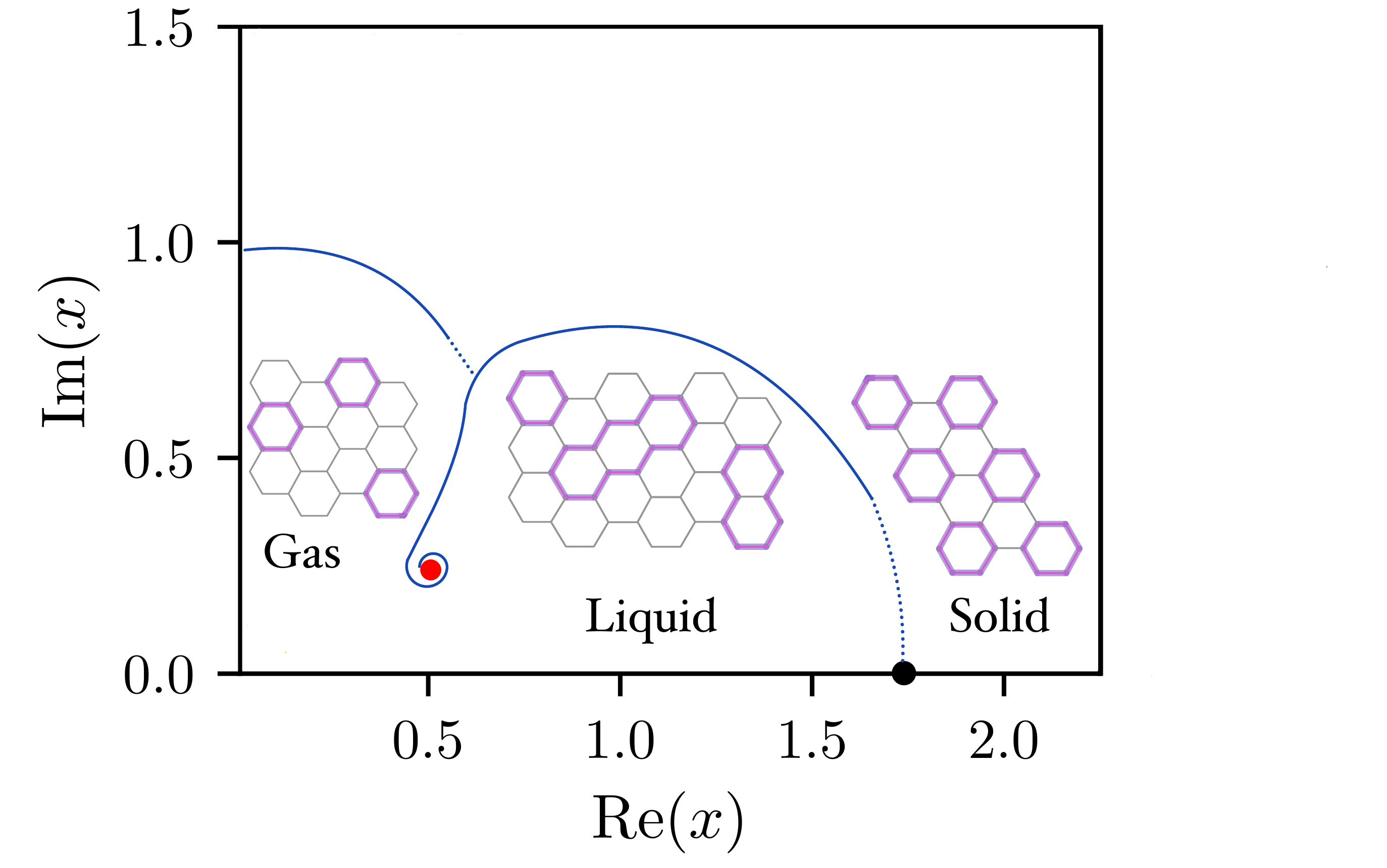} %
    \caption{Top: Scaled gap $\mathrm{Re}(X_t) \equiv (2\pi/L)\log(|\lambda_0|/|\lambda_1|)$ for $n=8$ and $L=9$. Note that we restricted the calculation of eigenvalues to the translation-invariant sector. The deep blue lines indicate equimodular lines. One of them approaches the predicted position of the CCFT $x_c(n)$, shown as a red dot. The black dot is the location of the $Q=3$ Potts transition $x_P(n=8)$ from Ref.~\cite{nlargex}. Bottom: Phase diagram suggested by the top panel. The inserts show pictorial representations of the three phases.    } %
    \label{fig:PhaseDiag}%
\end{figure}

However, this beta function fails to capture several properties of the flow.
First of all, it predicts $\frac{d \delta x}{dl} = i \sqrt{\mu} \delta x + \mathcal{O}(\delta x^2)$, with $\delta x = x- x_c$ the distance from the fixed point, which means the flow is actually circular around the CCFT until higher order terms in $\delta x$ are included.
This is not the case for our model since the linearized flow close to the fixed point is given by the scaling dimension: $\frac{d \delta x}{dl} = (2 - X_t) \delta x + \mathcal{O}(\delta x^2)$, and $\mathrm{Re}(2-X_t) \neq 0$.

Second, for any $n$, there should be an attractive fixed point at $x=0$ corresponding to the short loop phase (high-$T$ phase of the $O(n)$ model). 
We thus expect most of the trajectories emanating from the CCFT to reach $x=0$.
However, since the flow around the CCFT is a spiral, there needs to be a separatrix separating the trajectories which pass to the right or the left of the CCFT (see Fig. \ref{fig:RG1}(d)).
These features can be reproduced with the following generalized beta function:
\begin{equation}
    \beta_\text{gen}\left(x\right)=x\left(-\mu-(x-x_0)^2\right)\left(-\mu-(x+x_0)^2\right)
    \label{new_beta}
\end{equation}
where we added fixed points in the left plane which should be there by $x \to -x$ symmetry since the number of loop strands is always even.

What is the origin of the separatrix shown in blue in Fig.~\ref{fig:RG1}? 
A way to study this is to look at the thermal gap $X_t$ in the complex $x$ plane (see Fig.~\ref{fig:PhaseDiag} (top)).
We find that a line of $\mathrm{Re}(X_t)=0$ approaches the CCFT as $L \to \infty$.
Such a line is called \emph{equimodular} since it corresponds to $|\lambda_0| = |\lambda_1|$, and it is expected to host a finite density of zeros of the partition function in the thermodynamic limit~\cite{(7)_hard_hexagon_magnetic_gap_relation}.
Following earlier work~\cite{Liu_2011,Kim2008}, we propose to identify this line of zeros with the separatrix of the RG flow.
This line should approach the fixed point as $L$ goes to infinity following the spiral RG flow. In the Supplemental Material(SM~\cite{[{See Supplemental Material at }][]supp}), we show a finite size study of this flow based on the magnetic gap $X_h$.
This scenario is reminiscent of the Lee-Yang edge singularity on the imaginary axis of the magnetic field for the Ising model at $T>T_c$.
There, a line of zeros on the imaginary axis ends at a finite imaginary magnetic field $i h_c$ with the non-unitary Lee-Yang CFT~\cite{PhysRevLett.40.1610,PhysRevLett.54.1354}.
Note that the Lee-Yang CFT sits on the imaginary axis and its line of zeros approaches the critical point as a straight line, whereas here the line of zeros should actually approach the CCFT following a spiral.

The equimodular line can also be understood as a first-order transition line since it corresponds to a crossing of the two dominant eigenvalues of the transfer matrix. 
Based on an inspection of the corresponding eigenvectors (see SM~\cite{[{See Supplemental Material at }][]supp}), we interpret this first-order line as a transition between a gas phase to the left and a liquid phase to the right.
Pictorially, see Fig.~\ref{fig:PhaseDiag} (bottom), the gas phase is described as a dilute gas of single-hexagon loops, whereas the liquid phase has a comparatively larger weight on longer loops.
In this context, the CCFT is thus interpreted as a liquid-gas critical point located at the end of a first-order transition line.

Based on Fig.~\ref{fig:RG1}(d), we expect that all trajectories emanating from the CCFT end up at $x=0$, except for the separatrix.
However, the question remains of where the separatrix ends.
Fig.~\ref{fig:PhaseDiag} strongly suggests that it connects to another critical point which was previously reported to emanate from the $x=\infty$ point at $n>2$ in the $O(n)$ loop model~\cite{nlargex}.
Indeed, at large $x$, another bifurcation of critical points was observed at $n=2$ in the $O(n)$ loop model: a repulsive fixed point at $x^{-1}=0$ for $n \leq 2$ (describing so-called fully packed loops~\cite{Blote1994}) splits at $n=2$ into a repulsive fixed point at x=$x_P(n)$ and an attractive gapped fixed point at $x^{-1}=0$ which corresponds to a hard hexagon solid with a three-fold breaking of translation invariance.
The critical point at $x=x_P(n)$ was studied numerically in Ref.~\cite{nlargex} and was found to be consistent with $Q=3$ Potts criticality when $n$ is sufficiently large.
Overall, this suggests a gas-liquid-solid phase diagram with a liquid-gas first order line ending in a CCFT, and a melting transition described by $Q=3$ Potts.
We note that our conjectured beta function could be verified through a numerical RG analysis~\cite{Liu_2011}.

\textit{Discussion}---
In conclusion, we have established numerically the presence of CCFTs in the $O(n)$ loop model for $2 < n < n_g$, with $n_g \approx 12.34$.
We have also proposed a phenomenological beta function which reproduces the main features of the model, including a line of zeros which approaches the CCFT as $L \to \infty$ and serves as a separatrix for the RG flow.
We propose that this line of zeros can be understood as a first-order line transition between a gas-like and a liquid-like phase. 
We hope our results motivate further work on CCFTs in other contexts, like for non-Hermitian Hamiltonians or ``strange correlators''~\cite{You14,Scaffidi16,Scaffidi17}.


Regarding the abrupt disappearance of the CCFTs at $n_g$, an interesting observation is that $\mathrm{Arg}[x_{c,\pm}(n_g)] \simeq \mp \pi/6$. In the large-$n$ limit, typical configurations are dominated by the shortest loops, which are hexagons of length 6. The fugacity of these loops is $n x^6$ and the partition function thus becomes real for $\mathrm{Arg}[x]=\pm \pi/6$, which would explain why a CCFT cannot exist at that angle of the complex $x$ plane.
This argument is however only strictly correct in the large-$n$ limit and its extension to finite $n$ is left for future work.



A final point of discussion is the relation between the original $O(n)$ model and its loop formulation. 
The main difference between the two is that the former allows for loop crossings~\cite{Jacobsen03}. Loop crossings correspond to $4-$leg watermelon operators with scaling dimension $X_{l=4} = 3g/8 - 2/g+1$, which are irrelevant (i.e. $\mathrm{Re}(X_{l=4}) > 2$) for all $n>2$, so the CCFTs should exist in the original $O(n)$ model as well.


\begin{acknowledgements}
We would like to thank Adam Nahum for invaluable contributions to the manuscript and Xiangyu Cao for illuminating discussions.
T. S. acknowledges the support of the
Natural Sciences and Engineering Research Council of
Canada (NSERC), in particular the Discovery Grant
(No. RGPIN-2020-05842), the Accelerator Supplement
(No. RGPAS-2020-00060) and the Discovery Launch
Supplement (No. DGECR-2020-00222). This research
was enabled in part by support provided by Compute
Canada. 
Research at Perimeter Institute is supported in part by the Government of Canada through the Department of Innovation, Science and Economic Development Canada and by the Province of Ontario through the Ministry of Colleges and Universities.
\end{acknowledgements}

\bibliography{ref}

\appendix
\onecolumngrid
\setcounter{secnumdepth}{3}
\setcounter{figure}{0}
\setcounter{equation}{0}
\newpage\clearpage
\setlength{\belowcaptionskip}{0pt}

\begin{center}
    \large\textbf{Supplemental material to ``\textit{\titlename}''}
\end{center}

\section{Numerical determination of \texorpdfstring{$n_g$}{Lg}}
Numerically, we observe a level crossing at $n_g$ between the CFT ground state another state which appears gapped.

The value of $n_g$ has a finite-size scaling which can easily be deduced from the following argument.
Denoting $f_{CFT}(L,n)$ the free energy of the CFT ground state and $f_{Gapped}(L,n)$ the free energy of the competing gapped state, we identify $n_g(L)$ as the value of $n$ for which
\bea
\mathrm{Re}(f_{CFT}(L,n_g)) = \mathrm{Re}(f_{Gapped}(L,n_g))
\eea
We will leave the real part of all further equations implicit to ease notation.

For large enough $L$, the gapped state should only have exponentially small deviations from the thermodynamic limit: $f_{Gapped}(L,n) \simeq f_{Gapped}(L=\infty,n)$.
The CFT state, on the other hand, has the usual scaling: $f_{CFT}(L,n) \simeq f_{CFT}(L=\infty,n) + \frac{\pi c}{6 L^2}$.

This leads to
\bea
f_{CFT}(L=\infty,n_g(L)) + \frac{\pi c(n_g(L))}{6 L^2} = f_{Gapped}(L=\infty,n_g(L))
\eea
Now, using the Taylor expansion:
\bea
f(L=\infty,n_g(L)) \simeq f(L=\infty,n_g(L=\infty))+\left(\frac{\partial f(L=\infty,n)}{\partial n}\right)_{n=n_g(L=\infty)} (n_g(L) - n_g(L=\infty) )
\eea
for both $f_{CFT}$ and $f_{Gapped}$ and the expansion $c(n_g(L)) \simeq c(n_g(L=\infty))$ for the central charge, 
we find
\bea
&f_{CFT}(L=\infty,n_g(L=\infty))+\left(\frac{\partial f_{CFT}(L=\infty,n)}{\partial n}\right)_{n=n_g(L=\infty)} (n_g(L) - n_g(L=\infty) )  + \frac{\pi c(n_g(L=\infty))}{6 L^2} \\
&= f_{Gapped}(L=\infty,n_g(L=\infty))+\left(\frac{\partial f_{Gapped}(L=\infty,n)}{\partial n}\right)_{n=n_g(L=\infty)} (n_g(L) - n_g(L=\infty) )
\eea
Using $f_{Gapped}(L=\infty,n_g(L=\infty)) = f_{CFT}(L=\infty,n_g(L=\infty))$, we finally find
\bea
&\left(\frac{\partial f_{CFT}(L=\infty,n)}{\partial n}\right)_{n=n_g(L=\infty)} (n_g(L) - n_g(L=\infty) )  + \frac{\pi c(n_g(L=\infty))}{6 L^2} \\
&= \left(\frac{\partial f_{Gapped}(L=\infty,n)}{\partial n}\right)_{n=n_g(L=\infty)} (n_g(L) - n_g(L=\infty) )
\eea
which, after simplification, leads to
\bea
(n_g(L) - n_g(L=\infty) ) &= \frac{\pi c(n_g(L=\infty))}{6 L^2} \left(  \left(\frac{\partial f_{Gapped}(L=\infty,n)}{\partial n}\right)_{n=n_g(L=\infty)} - \left(\frac{\partial f_{CFT}(L=\infty,n)}{\partial n}\right)_{n=n_g(L=\infty)}  \right)^{-1} \\
&\equiv \frac{\alpha}{L^2}
\eea

Numerically, we can see that
\bea
\left(\frac{\partial f_{Gapped}(L,n)}{\partial n}\right) > \left(\frac{\partial f_{CFT}(L,n)}{\partial n}\right)
\eea
for the relevant range of $n$ around $n_g$, which means that $\alpha>0$, i.e. that $n_g(L)$ approaches $n_g(L=\infty)$ from above when $L$ increases.

\begin{figure}%
    \includegraphics[width=0.9\textwidth]{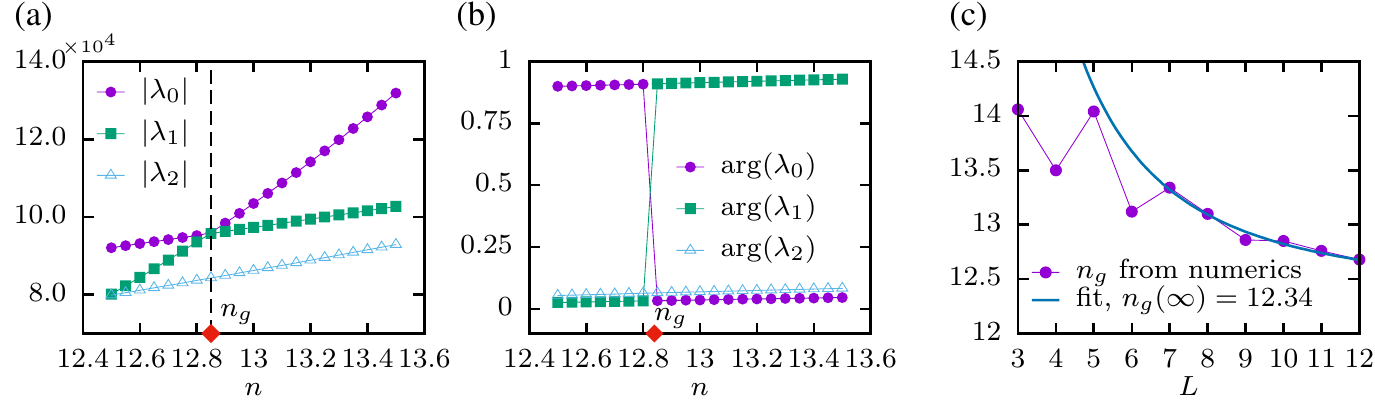} 
    \caption{Breakdown of CCFT: (a)The magnitude of the largest three eigenvalues of the transfer matrix (for $L=10$), showing a transition of the groundstate at $n=n_g$. (b) The phases of the largest three eigenvalues of the transfer matrix in radians. (c)$n_g$ vs $L$ and its fit with the theoretical prediction $n_g(L)=n_g(\infty)+\alpha/L^2$}%
    \label{fig:CCFT_breakdown}%
\end{figure}

\section{Analysis of the magnetic gap}

The renormalization group flow is predicted to move away from the critical point following a spiral since 
\begin{align}
\frac{d \delta x}{d l} = (2 - X_t) \delta x + \mathcal{O}(\delta x^2) 
\label{FlowLocal}
\end{align}
with $\delta x = x - x_{c}$ and $l$ the logarithmic scale.
We can deduce that the dynamics of $|\delta x|$ is controlled by $2 - \mathrm{Re}(X_t)$ and is repulsive for all $n>2$, whereas the rotation of $\mathrm{Arg}(\delta x)$ is controlled by $\mathrm{Im}(X_t)$ and is clockwise (resp. counterclockwise) around $x_{c,-}$ (resp $x_{c,+}$).

In our finite-size numerics, we can obtain a proxy for the RG flow by tracking how a specific feature of the partition function moves in the complex $x$ plane as $L$ is increased.
Inspired by Ref.~\cite{(7)_hard_hexagon_magnetic_gap_relation}, we find it convenient to do so by tracking points $x_d$ satisfying the following equimodular condition between the leading eigenvalues in the normal and ``magnetic'' sector of the transfer matrix: $|\lambda_{L,0}(x_d)| = |\tilde\lambda_{L,0}(x_d)|$.
Equivalently, this corresponds to a zero of the real part of the magnetic exponent $\mathrm{Re}(X_h(x_d,L))=0$.
When $L$ is large, we expect $x_d(L)$ to approach $x_c$ along the RG flow lines, $x_d(L) - x_c \sim L^{-(2-X_t)}$. It should thus spiral around $x_c$ for $n>2$.
In Fig 4, we show how $x_d(L)$ evolves with $L$ for various values of $n$.
We can see a clear contrast between $n<2$ and $n>2$: in the former case, $x_d(L)$ approaches $x_c$ from above following a straight line, whereas in the latter case, $x_d(L)$ exhibits a bent trajectory which is suggestive of a spiral flow for larger $L$.
\begin{figure}[ht!]
    \centering
    \includegraphics[width=0.48\textwidth]{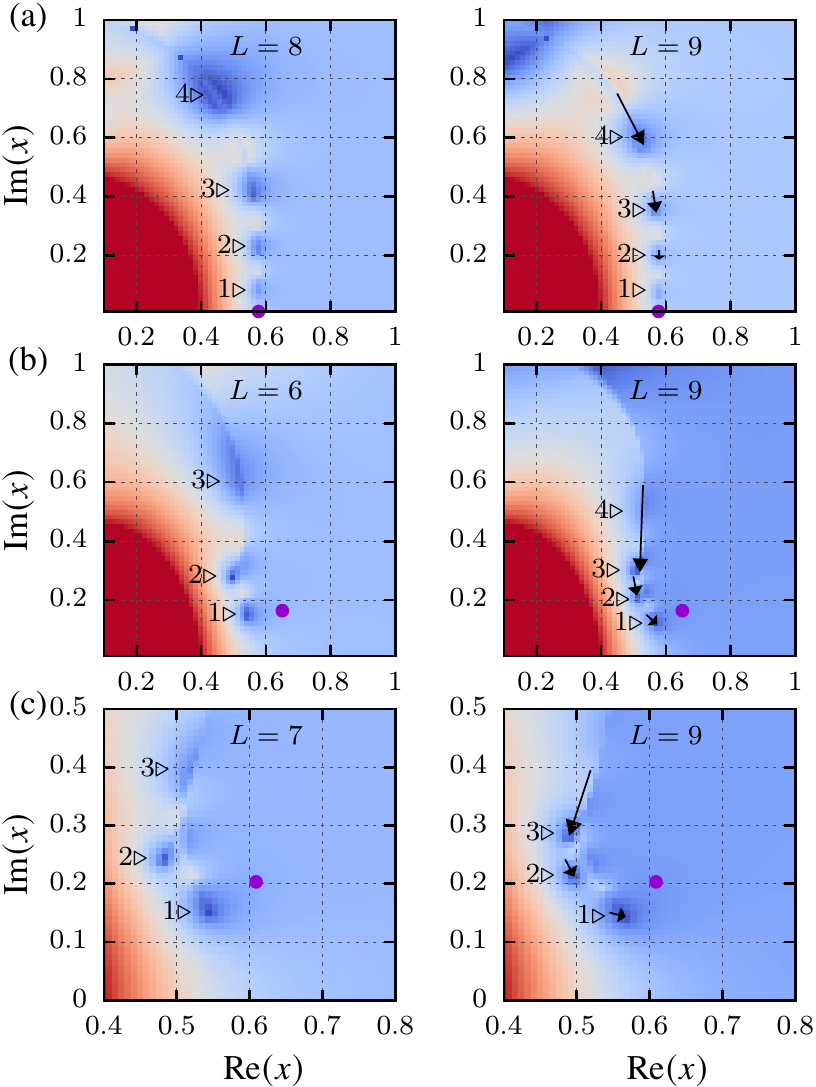}
    \caption{Evolution of the real part of the magnetic exponent $\mathrm{Re}(X_h)$ in the complex $x$-plane for $L=L_-$ (left) and $L=L_+$ (right). The loop fugacity is (a) $n=1.0$, (b) $ n=3.0$ and (c) $n=4.0$. A number of finite-size zeros of the magnetic gap (called $x_d(L)$ in the text) are indicated with 
    the numerals 1,2,\dots. Their flow towards the critical point $x_{c}$ (indicated with a filled circle) as $L$ increases is shown by the arrows in the right column, which point from $x_d(L_-)$ to $x_d(L_+)$.}
    \label{fig:RG}
\end{figure}

\section{Transfer Matrix Construction}
The lattice $O(n)$ loop model we study lives on an infinitely long cylinder having $L$ sites on the circumference. See \fig{fig:model} in the main text for a representation of the lattice where the cylindrical geometry is imposed using periodic boundary conditions (PBC) along the horizontal direction. The partition function for this cylindrical geometry is given in \eq{Z2} (main text), which we restate for convenience
\bea
Z = \sum_{i \in \text{loop config}} n^{N_i} x^{l_i}.
\label{app_eq:Z2}
\eea
\noindent
The transfer matrix $T$ provides a way to calculate the partition function $Z$ for the infinitely long cylinder using loop configurations defined on a cylindrical lattice of finite length $M$. The transfer matrix effectively adds a row of lattice sites, one layer at a time, onto the top edge of the finite cylinder, thus building up towards the infinite cylinder from bottom to top.\\
\begin{figure}
    \centering
    \includegraphics{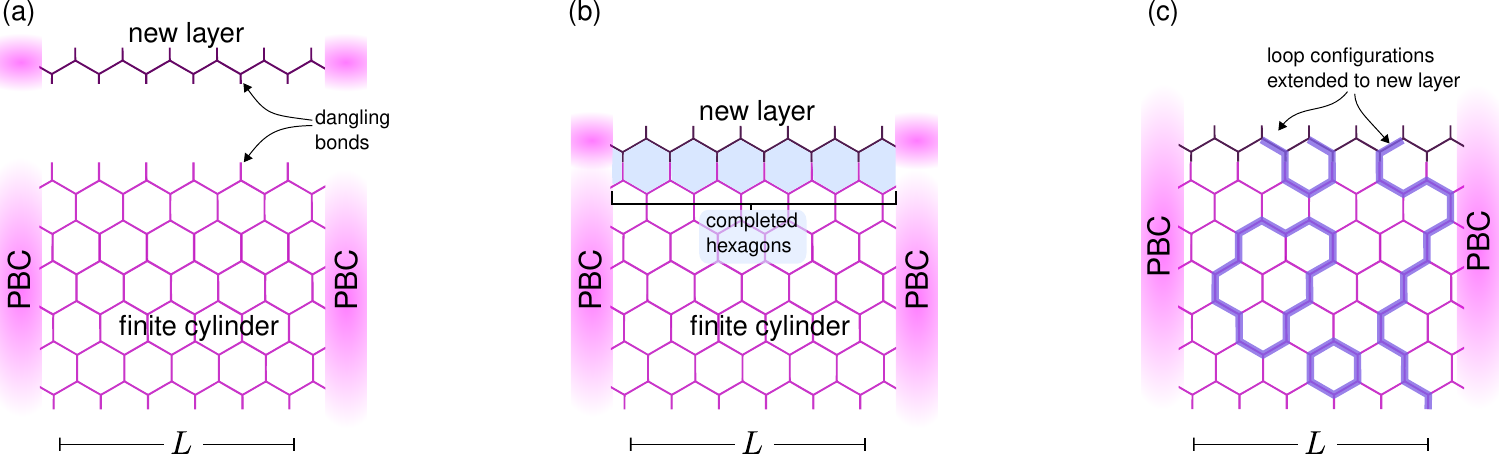}
    \caption{Construction of the transfer matrix. (a) The transfer matrix adds a new layer of lattice sites with dangling bonds to the top layer of the finite cylindrical lattice. (b) The dangling bonds from the new layer attach to dangling bonds of the existing cylinder to complete the hexagons of the honeycomb lattice. The length of the cylinder increases by one layer in the process. (c) Loop configurations appearing on the existing cylinder gets extended or propagated upward to the new layer by the application of the transfer matrix.}
    \label{fig:tmat_explanation}
\end{figure}
\noindent
For this purpose, it is helpful to visualize the $L$ lattice sites residing on the top edge of the finite cylinder with partially drawn vertically dangling bonds (see \fig{fig:tmat_explanation}(a)). Similarly, the lattice sites in the new layer to be added will also carry partial dangling bonds with them. In this way, when a new layer is added to the edge of the finite cylinder, the dangling bonds of the new lattice sites will connect to the existing partial dangling bonds to complete the hexagons of the honeycomb lattice (see \fig{fig:tmat_explanation}(b)).\\

\noindent
The state of a row of dangling bonds, whether belonging to existing or new lattice sites, is described by non-intersecting loop configurations. Specifically, each dangling bond can be occupied or unoccupied depending on whether a loop passes over it. Two occupied bonds are considered to be \emph{matched} if they are connected by the same loop occurring in the previously added layers. An occupied bond can be \emph{unmatched} or standalone if the bond is visited by a non-contractible loop. An example of a non-contractible loop is a loop running along the length of the infinitely long cylinder extending from one open boundary to another.\\

\noindent
We can represent the above scenarios using the following notation: a dot ``.'' for an unoccupied bond, a left ``\{'' or right ``\}'' brace for an occupied matched bond, and a vertical line ``$|$" for an occupied unmatched bond. Therefore, the collective state of a row of $L$ dangling bonds, which we call connectivity, is represented by a length $L$ string comprising the four symbols ``.'', ``\{'', ``\}'', ``$|$''. However, not all $4^L$ string possibilities are allowed. Since only non-intersecting loop configurations are permitted, the string denoting a connectivity $\alpha$ must be a balanced braces expression, where all opening and closing braces are correctly matched and nested while maintaining PBC. For example, the connectivity ``{}.{{.}}'' has three matched pairs of occupied bonds: one which connects bond 1 with bond 2, one which connects bond 4 with bond 8, and one which connects bond 5 with bond 7. Bonds 3 and 6 are unoccupied.\\

\noindent
Defining connectivity allows us to identify a ``conditional'' partition function $Z^{(M)}_\alpha$ \cite{Nienhuis82,Peled2017} for the finite cylinder of length $M$, which sums over loop configurations matching the connectivity $\alpha$ for the top dangling bonds, i.e.,
\bea
Z^{(M)}_\alpha = \sum_{i|\alpha} n^{N_i} x^{l_i}.\label{app_eq:Zma}
\eea
Using $Z^{(M)}_\alpha$, the role of the transfer matrix $\mathbf{T}$ can be stated precisely. The transfer matrix obtains the partition function of the cylinder with the added layer as a weighted sum over $Z^{(M)}_\alpha$\cite{Nienhuis82,Peled2017}
\bea
Z^{(M+1)}_\beta=\sum_\alpha \mathbf{T}_{\beta\alpha} Z^{(M)}_\alpha.\label{app_eq:Zma_Tba}
\eea
Here, $\alpha$ and $\beta$ are elements from the set of connectivities, each represented by balanced braces expression as discussed earlier. The element $\mathbf{T}_{\beta\alpha}$ of the transfer matrix, indexed by the connectivities $\alpha$ and $\beta$, is given by
\bea
\mathbf{T}_{\beta\alpha} = \sum_{m\in\{\alpha\to\beta\}} n^{n_l} x^{2L-n_v},\label{app_eq:Tba}
\eea
where $m$ iterates over all possible ways the loop configurations in connectivity $\alpha$ can be extended to reach configurations in $\beta$ when adding a new layer (see \fig{fig:tmat_explanation}(c)). While performing the actual computation of $\mathbf{T}_{\beta\alpha}$, we were able to list out these ways using a combinatorial approach of applying a set of finite moves to the connectivity $\alpha$ that allows us to reach $\beta$. The remaining symbols in \eq{app_eq:Tba} denote the number of empty (or unoccupied) bonds added $n_v$ and the number of loops closed $n_l$ by appending the new layer.\\

\noindent
In the asymptotic limit $M\to\infty$, the partition function $Z$ (\eq{app_eq:Z2}) can be related to $Z^{(M)}_\alpha$ (\eq{app_eq:Zma}) by appropriately tracing over the connectivities $\alpha$, and as a consequence of \eq{app_eq:Zma_Tba}, the partition function $Z$ is determined by the spectrum of $\mathbf{T}$. Using \eq{app_eq:Tba} and generating the table of all possible connectivities for a layer with $L$ sites, we compute and populate only the non-zero elements $\mathbf{T}_{\beta\alpha}$ of the transfer matrix. We diagonalize the final matrix using standard linear algebra subroutines provided by packages such as ARPACK to obtain the first few dominant eigenvalues of the transfer matrix.

\section{Characterization of the gas and liquid phases}
In the main text, we give an interpretation for the equimodular line approaching the CCFT as a line of first-order transition between a gas phase on the left and a liquid phase on the right.
In this appendix, we give evidence for this interpretation based on an observable which is a proxy for the average loop length, and which jumps abruptly from a small value in the gas phase to a larger value in the liquid phase.

\emph{Average loop length}---
We calculate $D$, which is a proxy for the average loop length, and is defined as:
\bea
D = \frac{\sum'_{\alpha } |\psi_{\alpha}|^2 D(\alpha)}{\sum'_{\alpha} |\psi_{\alpha}|^2 }
\eea
where $\sum'_\alpha$ is a sum over all connectivities except the empty one (i.e. except the one for which all bonds are inoccupied), where $\psi(\alpha)$ is the right eigenvector of the transfer matrix corresponding to the eigenvalue with largest magnitude $\lambda_0$, and where $D(\alpha)$ is the average distance between connected loop strands in connectivity $\alpha$. 
For example, for the connectivity $\alpha=\{\}.\{\{.\}\}$, the distance between matched braces is 1 for the pair connecting bonds 1 and 2, the distance is 4 for the pair connecting bonds 4 and 8, and the distance is 2 for the pair connecting bonds 5 and 7. This leads to an average of $D(\alpha) =  (1+4+2) / 3 = 7/3$.
Note that because we use periodic boundary conditions, the distance between bonds $i$ and $j$ with $i<j$ is defined as $\mathrm{min}(j-i,|j-i-L|)$.

In the gas phase, we expect to be dominated by the shortest loops and thus by connectivities of the type $\{\}..\{\}.\{\}.$.  This predicts a value of $D$ close to 1.
In the liquid phase, we expect to have $D$ substantially larger than 1 since there should be a comparatively larger weight on longer loops, and thus on connectivities with larger distances, like $\{...\}.\{..\}$. 

\begin{figure}[ht!]
    \centering
    \includegraphics[width=0.48\textwidth]{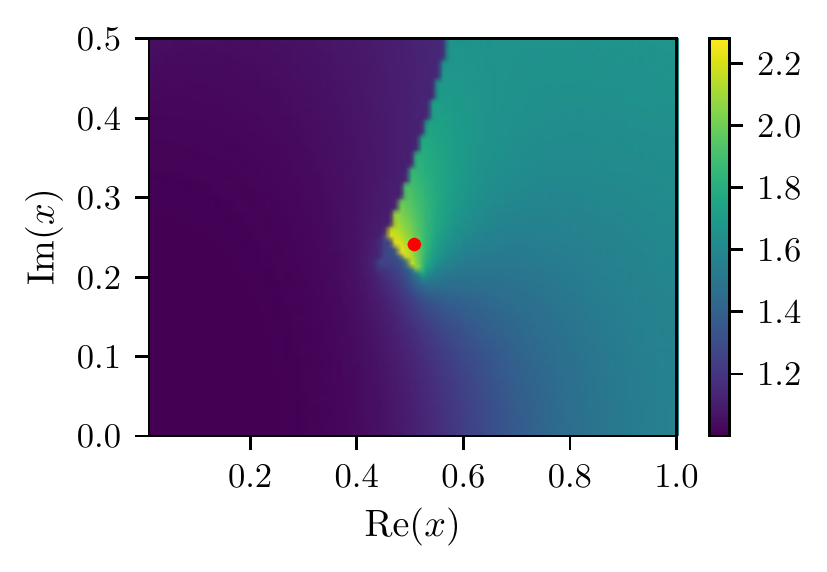}
    \caption{Map of average distance between connected bonds ($D$) for $n=8$ and $L=9$. The gas (left) and liquid (right) phases are separated by a first-order transition line which approaches the CCFT from above (the red dot shown the prediction for the CCFT location). Below the CCFT, the gas and liquid phases are smoothly connected, in analogy with a supercritical fluid.  }
    \label{fig:looplength}
\end{figure}

This is confirmed by Fig.~\ref{fig:looplength}, which shows that $D \simeq 1$ in the gas phase, and that $D$ jumps abruptly to a larger value when crossing the first-order transition line located above the CCFT.
On the other hand, when going from left to right below the CCFT, the value of $D$ is seen to change continuously in analogy with a supercritical fluid.

\end{document}